\documentclass[3p,times]{elsarticle}

\usepackage{ecrc}

\volume{00}

\firstpage{1}

\journalname{Nuclear Physics A}

\runauth{}

\jid{nupha}

\usepackage{amssymb}

\usepackage[figuresright]{rotating}

\newcommand{\jpsi}{$J/\psi$}
\newcommand{\pt}{$p_{ {\mathrm T} }$}
\newcommand{\y}{$y$}
\newcommand{\Raa}{$R_{\mathrm{AA}}$}
\newcommand{\Npart}{$N_{\mathrm{part}}$}
\newcommand{\gevc}{GeV/$c$}
\newcommand{\vtwo}{v$_{2}$}

\begin{document}

\begin{frontmatter}



\dochead{}

\title{Charmonia production in ALICE}


\author{Christophe Suire\corref{cor1}}
\author{for the ALICE Collaboration}
\address{Institut de Physique Nucl\'eaire d'Orsay, CNRS-IN2P3, Universit\'e Paris-Sud, France}
\cortext[cor1]{Email: christophe.suire@ipno.in2p3.fr}


\begin{abstract}
Quarkonia states are expected to provide essential information on the properties of the high-density strongly-interacting system formed in the early stages of high-energy heavy-ion collisions. 
ALICE is the LHC experiment dedicated to the study of nucleus-nucleus collisions and can  study  charmonia  at forward rapidity ($2.5 < y < 4$) via the $\mu^+ \mu^-$ decay channel and at mid rapidity ($|y| < 0.9$) via the $e^+ e^-$ decay channel.
In both cases charmonia are measured  down to zero transverse momentum.
The inclusive \jpsi\ production as a function of transverse momentum and rapidity in pp collisions at  $\sqrt{s} = $  2.76 and 7 TeV are presented. 
For pp collisions at  $\sqrt{s} = $  7 TeV, the inclusive \jpsi\  production as a function of the charged particle multiplicity, the inclusive \jpsi\  polarization at forward rapidity and the \jpsi\  prompt to non-prompt fraction are discussed.

Finally, the analysis of the inclusive \jpsi\ production in the Pb-Pb data  collected fall 2011 at a center of mass energy of $\sqrt{s_{NN}} = 2.76$ TeV is presented. 
Results on the nuclear modification factor are  then shown as a function of centrality, transverse momentum and rapidity and compared to model predictions.
First results on inclusive \jpsi\ elliptic flow are given.
\end{abstract}

\begin{keyword}
Hadron-induced high- and super-high-energy interactions, Relativistic heavy-ion collisions, Quark-Gluon plasma, \jpsi\ production and suppression mechanisms. 
\end{keyword}

\end{frontmatter}



\section{Charmonia in heavy-ion collisions}
\label{sec:intro}

Charmonia suppression via color-screening of the heavy-quark potential  was originally proposed as a probe of the QCD matter formed in relativistic heavy-ion collisions in 1986~\cite{Matsui:1986dk}.
\jpsi\ production  was extensively studied at the Super Proton Synchrotron (SPS) and at the Relativistic Heavy Ion Collider (RHIC). 
Indeed, \jpsi\ suppression in most central heavy-ion collisions was observed over a large energy range  ($\approx$ 20 to 200 \gevc ).    
The LHC has opened a new energy regime for the study of quarkonium in heavy-ion collisions. 
In a Pb-Pb collision at $\sqrt{s_{NN}} = 2.76$ TeV,  an average of  one \jpsi\  particle is expected to be produced in every central Pb-Pb collision, together with about 100 c$\bar{\rm{c}}$ pairs. 
Several models~\cite{{BraunMunzinger:2000px},{Thews:2000rj},{Andronic:2007bi},{Zhao:2007hh}} have included, already at RHIC energy,  a \jpsi\ regeneration component from deconfined charm quarks in the medium which counteracts the J/$\psi$  suppression in a QGP.
At LHC,  this regeneration component may become important, even dominant.

The in-medium modification of the \jpsi\ production can be  quantified with the nuclear modification factor \Raa , defined as the \jpsi\ yield measured in nucleus-nucleus collisions divided by the yield measured in pp collisions and the number of binary nucleon-nucleon collisions occurring in the nucleus-nucleus collision.
To interpret the \Raa , one must keep in mind the following points.   
First,  prompt \jpsi\ production in hadronic interactions consists of the sum of direct \jpsi\ ($\approx$ 65\%) and  excited $\rm{c \overline c}$ states such as $\chi_{\rm{c}}$ and $\psi\rm{(2S)}$ ($\approx$ 35\%). Since the  $\chi_{\rm{c}}$ and $\psi\rm{(2S)}$ have lower dissociation temperatures than the \jpsi , a \jpsi\ \Raa\ measurement around 0.65 is  compatible within errors with the suppression of excited states only. 
In addition to these prompt \jpsi\, one should also take into account that a non-prompt component from beauty hadron decays is present at LHC energy. 
The  \Raa\ includes   cold nuclear matter (CNM) effects, dominated by  nuclear absorption and (anti-) shadowing.  
These CNM effects can be responsible for  \jpsi\ suppression independently from  the creation of a deconfined medium.   
To quantify CNM effects,  proton-nucleus collisions are needed.  
At SPS energy, the  observed \jpsi\ suppression would be  compatible with the dissociation of excited states, once the CNM effects are taken into account.
At RHIC, a \jpsi\ \Raa\ of $\approx$ 0.25 in most central Au-Au collisions was measured by the  PHENIX  experiment~\cite{Adare:2006ns} with a strong centrality dependence. 
After estimating the correction  due to the CNM effects,  the suppression of direct \jpsi\ is at least $\approx$ 40\% or more.  
At the LHC,  \jpsi\  are  abundantly produced and  detailed studies of its production are possible in both elementary  and heavy-ion collisions, such as azimuthal asymmetry, polarization,  \Raa dependence on rapidity and on transverse momentum, etc.  
Such studies may give us some answers about the balance between suppression and recombination mechanisms of \jpsi .

The asymmetry the azimuthal distribution of \jpsi\  in the plane perpendicular to the beam direction, the so-called elliptic flow or \vtwo is indeed a  very interesting experimental observable.
When heavy-ions collide at finite impact parameter (non-central collisions), the geometrical overlap region and therefore the initial matter distribution is anisotropic and  
 is converted into a momentum anisotropy of the produced particles. 
The possible onset of  \jpsi\ production via recombination mechanisms should be, according to models, accompanied with a non-zero or possibly large \vtwo\ value~\cite{Liu:2009gx} at low \pt .  
Indeed,  if  charm quarks reach some level of thermalization in the medium, they  may acquire an elliptic flow that can be further transferred to the \jpsi\ assuming the \jpsi\ is formed via recombination. 
   
In the following section, the ALICE experiment and the data samples will be described.    
Then, \jpsi\  production results in pp collisions at $\sqrt{s} = $ 2.76 TeV and 7 TeV will be presented. 
The  \jpsi\  production in  Pb-Pb collisions at$\sqrt{s_{\mathrm{NN}} } = $ 2.76 will be studied through the \Raa\ dependence on centrality, \pt\ and \y .
Finally,  \jpsi\ elliptic flow  measurement will be shown.

\section{Experimental apparatus and data sample}
\label{sec:experiment}

ALICE is a general purpose  heavy-ion experiment and is described in~\cite{Aamodt:2008zz}. 
It consists of a central part covering the pseudo-rapidity  $ |\eta| < 0.9 $ and a muon spectrometer  covering   $ -4  <  \eta < -2.5$. 
At forward (mid) rapidity, \jpsi\ production is measured in the dimuon (dielectron) decay channel; in both cases the \pt\ coverage extends down to zero.  
Only detectors that are relevant to the analysis will be presented. 

At mid rapidity, the  \jpsi\ $\longrightarrow e^{+} e^{-}$ analysis makes use of the high precision tracking and particle identification of the Inner Tracking System  (ITS) and the Time Projection Chamber (TPC).
The ITS consists of six cylindrical layers of silicon detectors; at a radius 3.9 and 7.6 cm, the first two layers are equipped  with silicon pixels (SPD), then two layers of  silicon drifts  at radius 15 and 23.9 cm and finally, two layers of silicon strips at radius  38 and 43 cm. 
Its main tasks are the primary and secondary vertex reconstruction; the resolution on the primary vertex ranges from  100 $\mu$m (pp collisions) to 10 $\mu$m (central Pb-Pb collisions).      
The SPD has triggering capabilities and can provide a signal at level 0. 
The large cylindrical TPC has full azimuthal coverage and extends from z = -2.50 m to z = 2.50 m~\footnote{The z axis is defined here as the beam line axis in the counter clockwise direction and its origin is at the center of the ALICE detector.}.  
The TPC radial coverage ranges from r = 85 cm to r = 247 cm.
This  large drift detector is the main track reconstruction device at central rapidity since it  can provide up to 159 space points per track. 
Particle identification (PID) is achieved via the measurement of the specific energy loss (dE/dx) of particles in the detector gas (Ne/CO$_{2}$/N$_{2}$). 
The excellent dE/dx resolution of $\approx$ 5\%  allows to identify electrons by using inclusion cut  around the Bethe-Bloch  fit for electrons and exclusion cuts  for protons and pions. 
ALICE has further capabilities to improve electron identification and triggering thanks to the  Time-Of-Flight (TOF), the Transition Radiation Detector (TRD) and the Electromagnetic Calorimeter (EMCAL) detectors. However, these detectors have not  been  used in the analysis presented here. 

At forward rapidity  ($2.5 < $ \y\  $ < 4$) the production of quarkonium states is measured in the muon spectrometer~\footnote{In the ALICE reference frame,  the muon spectrometer covers a negative $\eta$ range and consequently a negative $y$ range. We have chosen to present our results with a positive $y$ notation.}. 
The spectrometer consists of a ten interaction length thick absorber ( -0.9 m $<$ z $<$ -5.0 m) filtering the  muons in front of five tracking stations (MCH) made of two planes of cathode pad chambers each. 
The third station is located  inside a dipole magnet with a 3 Tm field integral. 
The MCH chambers are positioned between z=-5.2 and z=-14.4 m.
The tracking apparatus is completed by a triggering system (MTR) made of two stations, located at z=-16.1 and z=-17.1 m, each equipped of two planes of resistive plate chambers. 
The MTR chambers are downstream of a 1.2 m thick iron wall,   which absorbs secondary hadrons escaping from the front absorber and low momentum  muons coming mainly from $\pi$ and K decays.
Throughout its full length, a conical absorber made of tungsten, lead and steel protects the muon spectrometer against secondary particles produced by the interaction of large-$\eta$ primaries in the beam pipe.
The forward VZERO detectors, two arrays of 32 scintillator tiles  covering the range $2.8 \leq \eta \leq 5.1$ (VZERO-A) and $-3.7 \leq \eta \leq -1.7$ (VZERO-C), are positioned at z=340 and z=-90 cm. 
And finally, the zero degree calorimeters (ZDC) placed 116 m down and up-stream ALICE can detect spectator neutrons and protons. 

In proton-proton collisions, the minimum bias (MB) trigger uses information of the SPD and VZERO detectors. It is defined as the logical OR of the three following conditions: 
(i) a signal in  two  readout chips in the outer layer of the SPD,  (ii) a signal in VZERO-A,  (iii) a signal in VZERO-C. 
This MB trigger requires the coincidence of the crossing of two proton bunches at the experiment interaction point (IP). 
ALICE MB trigger selects about 86\% of the proton-proton inelastic cross section. 
Specific cross sections were measured during van der Meer scan in pp collisions at 7 and 2.76 TeV  and allowed to determine  the absolute normalization of the inclusive \jpsi\ cross section.
The muon minimum bias trigger ($\mu$-MB) requires, in addition to the MB conditions given above, a signal in the MTR system.  
The MTR can reconstruct a {\it trigger track},  determine its \pt\ and select different thresholds (\pt\ $\approx$ 0.5, 1 and 4 \gevc ).
The $\mu$-MB trigger helps to take  advantage of the full luminosity delivered by the LHC  in the muon spectrometer. 
The MB trigger used in Pb-Pb collisions collected in 2010  requires the logical AND of the conditions (i),(ii) and (iii) given above.   
The centrality of the collision is determined from the amplitude of the VZERO signal fitted with a geometrical-Glauber model~\cite{PhysRevLett.106.032301}. 
In 2011, the MB conditions were reduced to  the AND of conditions (ii) and (iii) but additional requirements were added to select rare events. 
In particular, event multiplicity and dimuon triggers were added and  ZDC were used for rejecting electromagnetic Pb-Pb interactions and satellite Pb-Pb collisions.
Once the centrality selection cut has been applied, triggers are fully efficient with negligible contamination.

\section{\jpsi\ production  in pp collisions}
\label{sec:ppresults}

The  \jpsi\  production in pp collisions is extensively studied in ALICE and only a selection of the available results will be presented in this section. 
Further details on the related analysis can be found  in~\cite{Geuna:hp2012}. 

The inclusive  \jpsi\  production was measured in pp collisions at $\sqrt{s} = $ 7 TeV in the dimuon and dielectron channels in the rapidity ranges $2.5 < y < 4 $ and $ |y| < 0.9$  down to \pt\ = 0. The analysis was made with an integrated luminosity $\mathcal{L}_{\rm int} \approx 16 \; (6) \;  \mathrm{nb}^{-1}$ in the dimuon (dielectron) channel. 
The measured cross section values are $\sigma_{J/\psi} (|y|<0.9) = 10.7 \pm 1.0 (\mathrm{stat.}) \pm 1.6 (\mathrm{syst.}) ^{+1.6}_{-2.3} (\mathrm{syst. pol.}) \;  \mu \mathrm{b} $ and $\sigma_{J/\psi} (2.5<y<4.) = 6.31 \pm 0.25(\mathrm{stat.}) \pm 0.76 (\mathrm{syst.})  ^{+0.95}_{-1.96} (\mathrm{syst. pol.})  \;  \mu \mathrm{b} $.
At forward rapidity differential cross section  d$^{2}\sigma$/d\pt d\y\ measurement from ALICE fully overlaps with LHCb and a good agreement is found.
At mid rapidity, the situation is different since ATLAS and CMS cannot measure \jpsi\ with \pt\ $\lesssim$ 6 \gevc . 
Thus combining ALICE and CMS/ATLAS data offers a rather complete inclusive  \jpsi\  production measurement over a large rapidity range.
These comparisons are available in~\cite{Aamodt:2011gj}.

The same analysis was carried out with pp collisions at $\sqrt{s} = $ 2.76 TeV collected in March 2011. 
Since the center of mass energy per nucleon-nucleon collisions is identical to the one of the Pb-Pb collisions, this analysis provides an essential reference data to measure the \jpsi\ nuclear modification factor. 
The integrated luminosity for the analysis is  $\mathcal{L}_{\rm int} \approx 20 \; (1) \;  \mathrm{nb}^{-1}$ in the dimuon (dielectron) channel. 
The integrated cross sections are $\sigma_{J/\psi} (|y|<0.9) = 6.71 \pm 1.24 (\mathrm{stat.}) \pm 1.22 (\mathrm{syst.}) ^{+1.01}_{-1.41} (\mathrm{syst. pol.})  \;  \mu \mathrm{b} $ and $\sigma_{J/\psi} (2.5<y<4.) = 3.34 \pm 0.13(\mathrm{stat.}) \pm 0.28 (\mathrm{syst.})  ^{+0.53}_{-1.07} (\mathrm{syst. pol.})  \;  \mu \mathrm{b} $.
Note that the uncertainties quoted  here  on the pp measurement are one of the main source of uncertainty of the nuclear modification factor discussed in the next section. 
The differential cross sections  d$^{2}\sigma$/d\pt d\y\ have been extracted down to \pt\ = 0 at both rapidities.  
These results are  compared to a theoretical model, NRQCD  calculation  that includes Color Singlet and Color Octet terms at NLO, which describes reasonably well the measurement at  $\sqrt{s} = $ 2.76 TeV and also the one at  $\sqrt{s} = $ 7 TeV~\cite{Aamodt:2011tmp}.

In the previous results, one could remark that the \jpsi\ cross section has a large uncertainty related to the unknown polarization.
ALICE has studied \jpsi\ polarization in pp collisions $\sqrt{s} = $ 7 TeV in the dimuon channel.
Measurements of the polar and azimuthal angle distributions  of the decay muons allowed us to extract the \jpsi\ polarization  for 2 $<$ \pt\ $<$ 8 \gevc\ and 2.5 $<$ \y\ $<$ 4.
The parameters describing the \jpsi\ polarizations are consistent with zero in the kinematic range under study~\cite{Abelev:2011md}.
This measurement is, at the present date, the only \jpsi\ polarization measurement at the LHC. 
It is crucial in the  near future to extend  the polarization measurement down to zero \pt\ and to high \pt\ in order to provide   more stringent tests to theoretical calculations.
In addition, since the pp cross section enters the nuclear modification factor calculation, the polarization, if different from zero,  may have a strong  impact at low transverse momentum.
Such a measurement needs a large statistic and strengthens the requirement to collect a large  amount of data at the same center of mass energy as the Pb-Pb collisions.

An interesting feature of the \jpsi\ production in pp collisions at  $\sqrt{s} = $ 7 TeV arises from its dependence on  the charged particle multiplicity. 
The d$N_{\rm{ch}}$/d$\eta$ is calculated from the number of tracks reconstructed in $|\eta| < 1 $ using pairs of hits (tracklets)  in the SPD.
These measurements were performed at both rapidities in the dimuon and dielectron channels.
Expressed in terms of the relative \jpsi\ yield     $\frac{\mathrm{d}N_{J/\psi}/\mathrm{d}\eta}{\langle  \mathrm{d}N_{J/\psi}/\mathrm{d}\eta \rangle}$ as a function of  the relative charged multiplicity  $\frac{\mathrm{d}N_{\mathrm{ch}}/\mathrm{d}\eta}{\langle  \mathrm{d}N_{\mathrm{ch}}/\mathrm{d}\eta \rangle}$, a linear increase is clearly seen at both rapidities. 
For $\mathrm{d}N_{\mathrm{ch}}/\mathrm{d}\eta / \langle  \mathrm{d}N_{\mathrm{ch}}/\mathrm{d}\eta \rangle  = 4 $ ($\approx 24/6$), the relative \jpsi\ yield is enhanced by a factor of about 5 at forward rapidity and about 8 at mid rapidity. 
This trend is not reproduced by PYTHIA 6.4.25 in the Perugia 2011 tune which exhibits an opposite tendency, i.e. a decrease of the \jpsi\ multiplicity with respect to the event multiplicity~\cite{Abelev:2012rz}.  
One could infer that the  \jpsi\ production is strongly connected with the underlying hadronic activity. 
Whether this hadronic activity comes from multiple parton interactions  remains an open question. 
Further investigations are needed to better understand this measurement that challenges our understanding of the \jpsi\ production in pp collisions. 
In particular, the event multiplicity dependence should be completed by the \pt\ dependence and extended to the open charm cross section (e.g. D mesons).

All the results presented up to now refer to inclusive \jpsi\ production which sums three distinct contributions:  the prompt \jpsi\ produced directly in the pp collisions, the prompt  \jpsi\ produced indirectly via the decay of heavier charmonia states and the non-prompt \jpsi\ produced in the decay of beauty hadrons.    
At central rapidity ($ |y| <  0.9$), the measurement of the non-prompt \jpsi\ was achieved in pp collisions at $\sqrt{s} = $ 7 TeV for $ 1.3 <$ \pt\ $< 10$ \gevc . 
This measurement is only accessible in ALICE since the other experiments cannot detect \jpsi\  at mid rapidity below a  \pt\ of 6.5  \gevc\ where most of the cross section lies. 
The integrated luminosity for the analysis is  $\mathcal{L}_{\rm int} = 5.6  \mathrm{nb}^{-1}$. 
This measurement relies on the discrimination of \jpsi\ produced detached from the primary vertex of the pp collisions thanks to the good spatial resolution of the ITS. 
By fitting simultaneously the invariant mass spectra and the pseudo-proper decay length of the reconstructed \jpsi , one can can measure the relative abundances of  prompt and non-prompt \jpsi , and the background. 
The fraction of \jpsi\ from beauty hadrons ($f_{\mathrm{B}}$) in the  measured kinematic range is about 15\% with a strong \pt\ dependence. 
Then, $f_{\mathrm{B}}$ is combined with the  the inclusive \jpsi\ cross section measured in ~\cite{Aamodt:2011gj} to extract the prompt \jpsi\ cross section 
 $\sigma_{J/\psi}^\mathrm{prompt} (|y|<0.9, p_{\mathrm{T}} > 1.3 {\mathrm{~GeV}/c}) = 7.2  \pm 0.7 (\mathrm{stat.}) \pm 1.0 (\mathrm{syst.}) ^{+1.3}_{-1.2} (\mathrm{syst. pol.})  \;  \mu \mathrm{b} $. 
Comparisons with models lead to a good description of the prompt \jpsi\ dependence with \pt\ and of the total beauty cross section~\cite{Abelev:2012gx}.

\section{Nuclear modification factor in Pb-Pb collisions at $\sqrt{s_{\mathrm{NN}} } = $ 2.76 TeV}
\label{sec:raaresults}

Inclusive  \jpsi\ production was studied in Pb-Pb collisions at 2.76 TeV  at mid and forward rapidity in the dielectron and dimuon decay channels using respectively an integrated luminosity  $\mathcal{L}_{\rm int} \approx $ 2.1 and  $\mathcal{L}_{\rm int} \approx 70 $ $\mu$b$^{-1}$.
A crucial feature of the ALICE detector is to measure, in both channels, the \jpsi\ production down to \pt\ = 0 GeV/c.
The large data sample analyzed in the dimuon channel allowed us to perform differential analysis of the nuclear modification factor (\Raa ) as function of centrality, \pt , and \y . 
In the dielectron channel, only the centrality dependence in 3 centrality classes (0--10\%, 10--40\% and 40--80\%) could be achieved. 
One should note that the acceptance times efficiency factor in the dielectron (dimuon) channel is quite high $\approx 8\%$ ($14\%$)  and weakly  depends on the collision centrality with a maximum relative loss of 12\% (8\%) from peripheral to most central collisions.    
Details on both analysis are given in~\cite{Wiechula:hp2012}. 
\begin{figure} [t!]
\begin{center}
\includegraphics[width=0.49\linewidth,keepaspectratio]{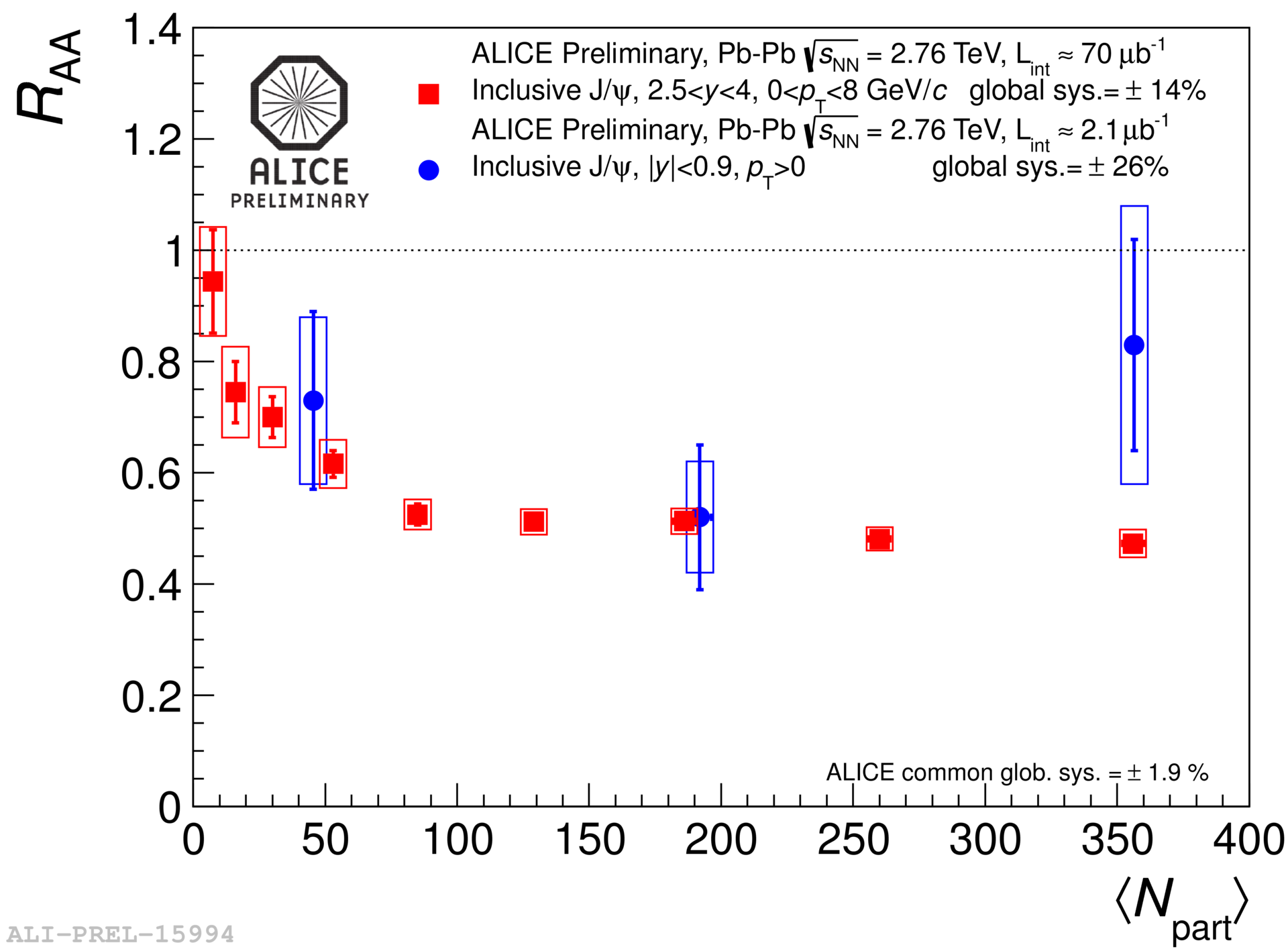}
\includegraphics[width=0.49\linewidth,keepaspectratio]{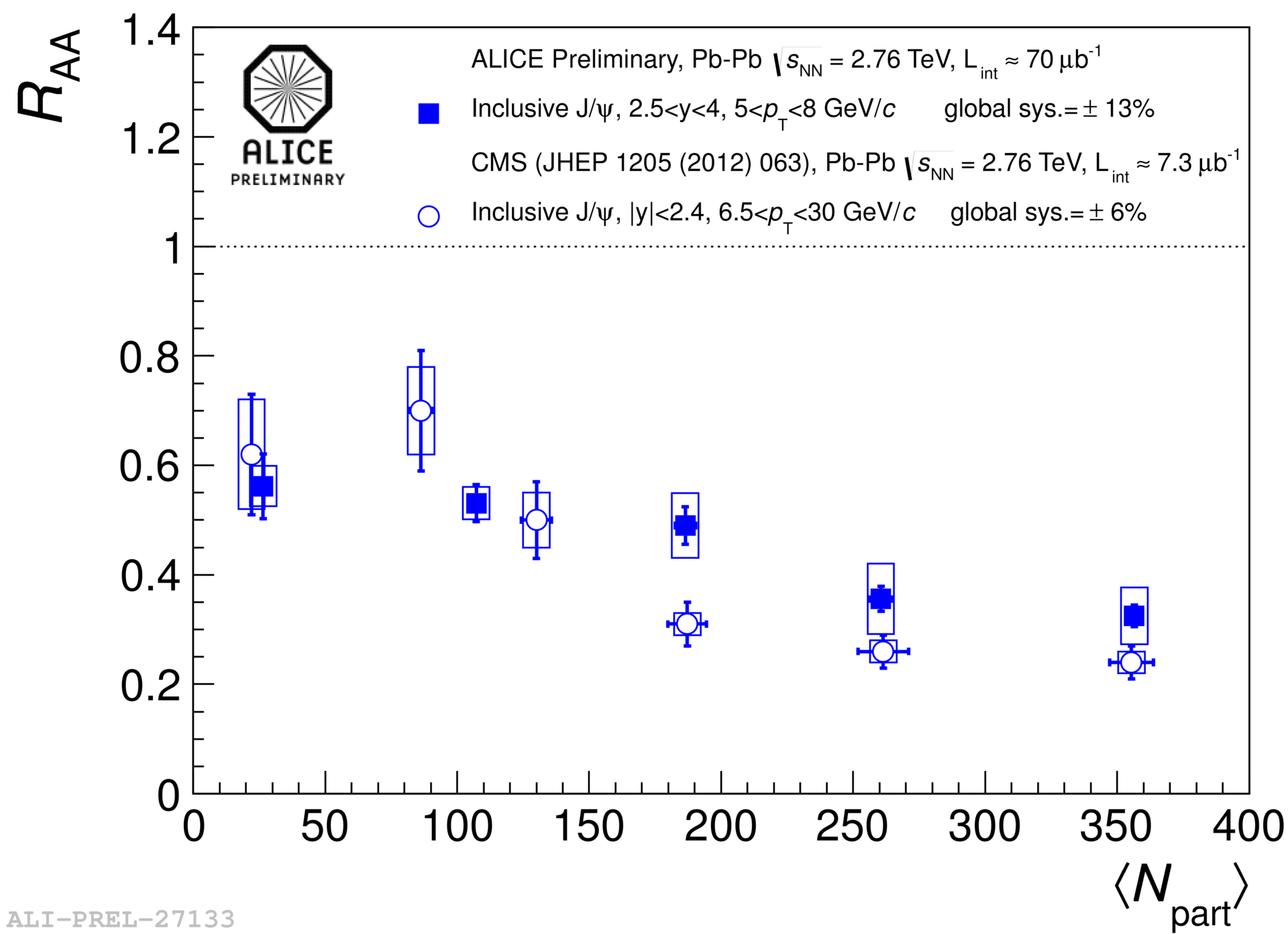}
\caption{(Color online) \label{fig:raaexp} Inclusive \jpsi\ $R_{\rm{AA}}$ measured in Pb-Pb collisions at $\sqrt{s_{\mathrm{NN}}} = 2.76$ TeV 
at forward and mid rapidity in ALICE is shown on the left  side.  
On the right side the \jpsi\ \Raa\ at high-\pt\  from ALICE at forward \y\ is compared to one measured by CMS at central \y .}
\end{center}
\end{figure}
On the left side of Fig.~\ref{fig:raaexp}, the inclusive  \jpsi\ \Raa\ is shown as a function of the number of nucleons participating in the collision~\cite{PhysRevLett.106.032301} at mid and forward rapidity. 
At forward rapidity, a clear suppression is seen for \Npart\ $ > 70 $ with  almost no centrality dependence. 
These results show a good agreement with the ones published 
in~\cite{PhysRevLett.109.072301}  
 based on $\mathcal{L}_{\rm int} \approx 2.9 $  $\mu$b$^{-1}$ collected in 2010.   
At mid rapidity, a similar pattern could be possible but the coarser centrality classes and larger uncertainties prevent to draw any firm conclusion.
The  centrality integrated \jpsi\ \Raa\  at forward and mid rapidity are $R^{0\%-90\%}_{\rm{AA}} = 0.497 \pm  0.006 \rm{(stat.)} \pm 0.078 \rm{(syst.)}$ and   $R^{0\%-80\%}_{\rm{AA}} = 0.66 \pm  0.10 \rm{(stat.)} \pm 0.24 \rm{(syst.)}$.
In both cases, the systematic uncertainty is dominated by the pp reference. 
On the right side of Fig.~\ref{fig:raaexp}, the centrality dependence of \jpsi\ \Raa\ at high-\pt\  is compared  between ALICE and CMS~\cite{Chatrchyan:2012np}.  
A larger suppression, \Raa\ $\approx 0.25-0.30$,  is measured in the most central collisions with a clear centrality dependence. 
One could see here an indication that the  \jpsi\ \Raa\  is \pt\ dependent at forward rapidity and possibly at mid rapidity; selecting high-\pt\  \jpsi\ drives down the \Raa.

\begin{figure}
\begin{center}
\includegraphics[width=0.49\linewidth,keepaspectratio]{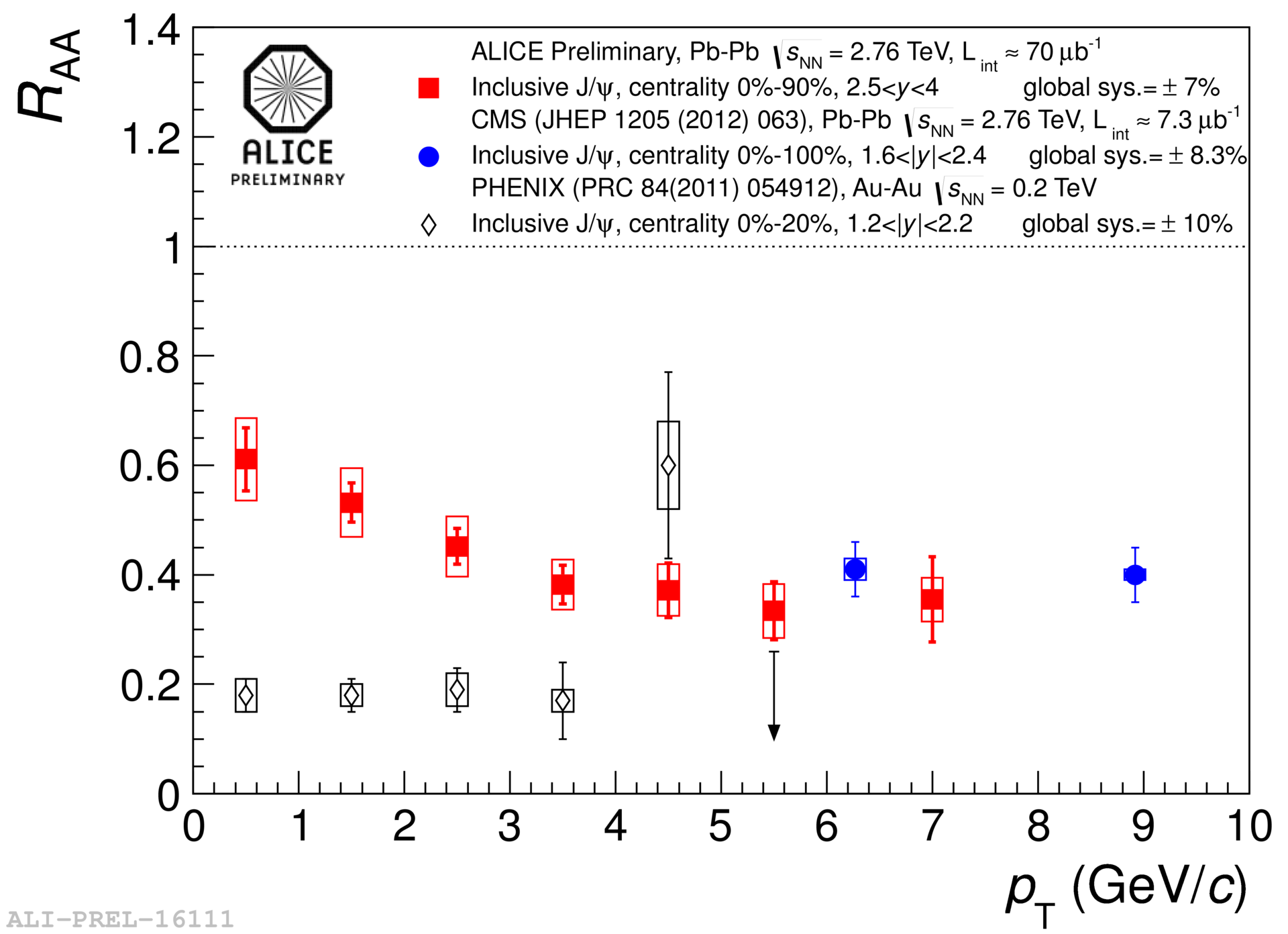}
\includegraphics[width=0.49\linewidth,keepaspectratio]{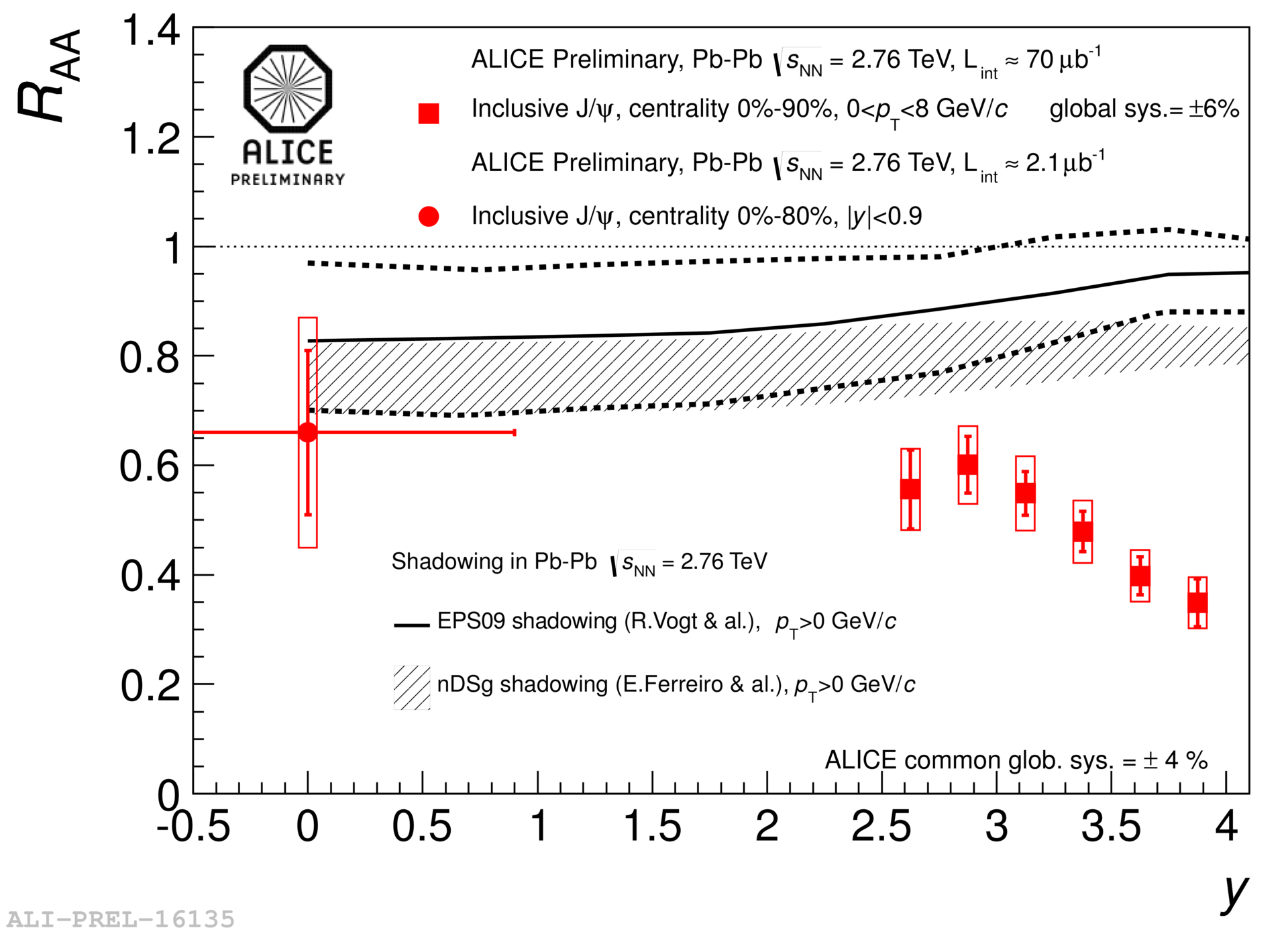}
\caption{(Color online) \label{fig:raaexp2} ALICE \pt\ dependence of the inclusive \jpsi\ \Raa\ measured in Pb-Pb collisions at $\sqrt{s_{\mathrm{NN}}} = 2.76$ TeV is compared to PHENIX and CMS measurements on the left side. On the right side, the rapidity dependence of the  \jpsi\ \Raa\ measured by ALICE  is compared with model predictions~\cite{Vogt:2010aa,Ferreiro:2011rw} that implement only shadowing effects.}
\end{center}
\end{figure}

The \pt\ dependence of the \jpsi\ \Raa\ is confirmed and can be better observed in Fig.~\ref{fig:raaexp2} (left side). 
The inclusive \jpsi\ \Raa\ is shown as a function of \pt\ for the 0\%--90\% most central Pb-Pb collisions and exhibits a decrease from 0.6 to 0.35 approximately.    
At high-\pt\, a rather direct comparison with CMS results~\cite{Chatrchyan:2012np} is possible; the only difference is that the CMS measurement covers a more central rapidity range ($ 1.6  < |y| < 2.4 $).
A reasonable agreement between the two measurements is found for high-\pt\ \jpsi\ \Raa .
For \pt\ smaller than 4 \gevc , the difference with PHENIX measurement~\cite{Adare:2011yf} is striking. 
The PHENIX result concern the 0\%--20\%  most central Au-Au collisions whereas the ALICE result is for a much wider centrality range (0\%--90\%). However, the bulk of the \jpsi\ production ($\approx 60 \%$) occurring in 0\%--20\% most central collisions, the comparison remains meaningful. 
In addition, work is ongoing to extract the \Raa\ versus \pt\ in smaller centrality classes. 

The \jpsi\ \Raa\ dependence on rapidity has been measured over a wide range thanks to the combination of our measurement in the central barrel and in the muon spectrometer, and is displayed on the right side of  Fig.~\ref{fig:raaexp2}.
At forward rapidity, the  \jpsi\ \Raa\ decreases by $\approx 40\%$ from $y=2.5$ to $y = 4$.  
The  measurement at mid rapidity, because of its large uncertainties,  does not allow to draw a clear conclusion but hints towards a rather flat behavior between \y\ = 2.5 and \y\ = 0.  
On the same figure, an estimate of the \jpsi\ \Raa\ due only to shadowing effects is given for two models. 
Indeed at LHC energies, modification of the gluon distribution  function is dominated  by shadowing effects~\cite{Lourenco:2008sk}.
The first model is a Next to Leading Order calculation within the Color Evaporation Model~\cite{Vogt:2010aa} with the EPS09 nuclear PDF (nPDF). 
The second model is a   Leading Order calculation within the CS  Model~\cite{Ferreiro:2011rw} with the nDSg nPDF.
In the first model,  the upper and lower limits correspond to the uncertainty of the EPS09 nPDF,  and in the second model the band covers the uncertainty in the factorization scale of the for nDSg PDF.
One could not exclude that shadowing effects are responsible  for a large part of the \jpsi\ suppression observed  in \Raa\ from \y\ = 0 to \y\  $\approx$ 3, this would imply that the expected color screening  \jpsi\ suppression observed at lower energy (RHIC)  or higher \pt\ (CMS) is either small  or compensated by recombination mechanisms.   
In the rapidity range from 3 to 4, our results show that the suppression goes beyond the shadowing-only  prediction  given  by models with our current knowledge of nPDF.

The influence of the contribution of beauty hadron feed-down to the inclusive \jpsi\ yield in our \y\ and \pt\ range was estimated.  
Non-prompt \jpsi\ are indeed different since their suppression or production  is  insensitive  to color screening or recombination phenomena that are expected to occur in 
the hot and dense medium created in the Pb-Pb collisions. 
The beauty hadron  decay mostly occurs  outside the fireball, and a measurement of the non-prompt \jpsi\ \Raa\ is connected to  the beauty quark in-medium energy loss.
At forward rapidity,  the non-prompt \jpsi\ was measured  by the LHCb collaboration to be about 10\%  in pp collisions at $\sqrt{s} = 7$ TeV~\cite{Aaij:2011jh} in our \pt\ range. 
Assuming  the scaling of beauty production  with the number of binary nucleon-nucleon collisions  and neglecting the shadowing effects, the prompt \jpsi\ \Raa\ would be, and this is an upper limit,  11\% smaller than our inclusive measurement. 
To estimate the influence of non-prompt \jpsi\ as a function of \pt\ and \y\ on our inclusive \Raa\ measurement, we have extrapolated the LHCb measurement at  $\sqrt{s} = 7$ TeV  down  $\sqrt{s} = 2.76$ TeV using an center of mass energy dependence extracted from  CDF and  CMS data.  
Assuming a range of energy loss for the beauty quarks from \Raa(b) = 0.2 to \Raa(b)= 1,  we  have found that the \jpsi\ from beauty hadrons have a negligible influence on our measurement.

\begin{figure}
\begin{center}
\includegraphics[width=0.49\linewidth,keepaspectratio]{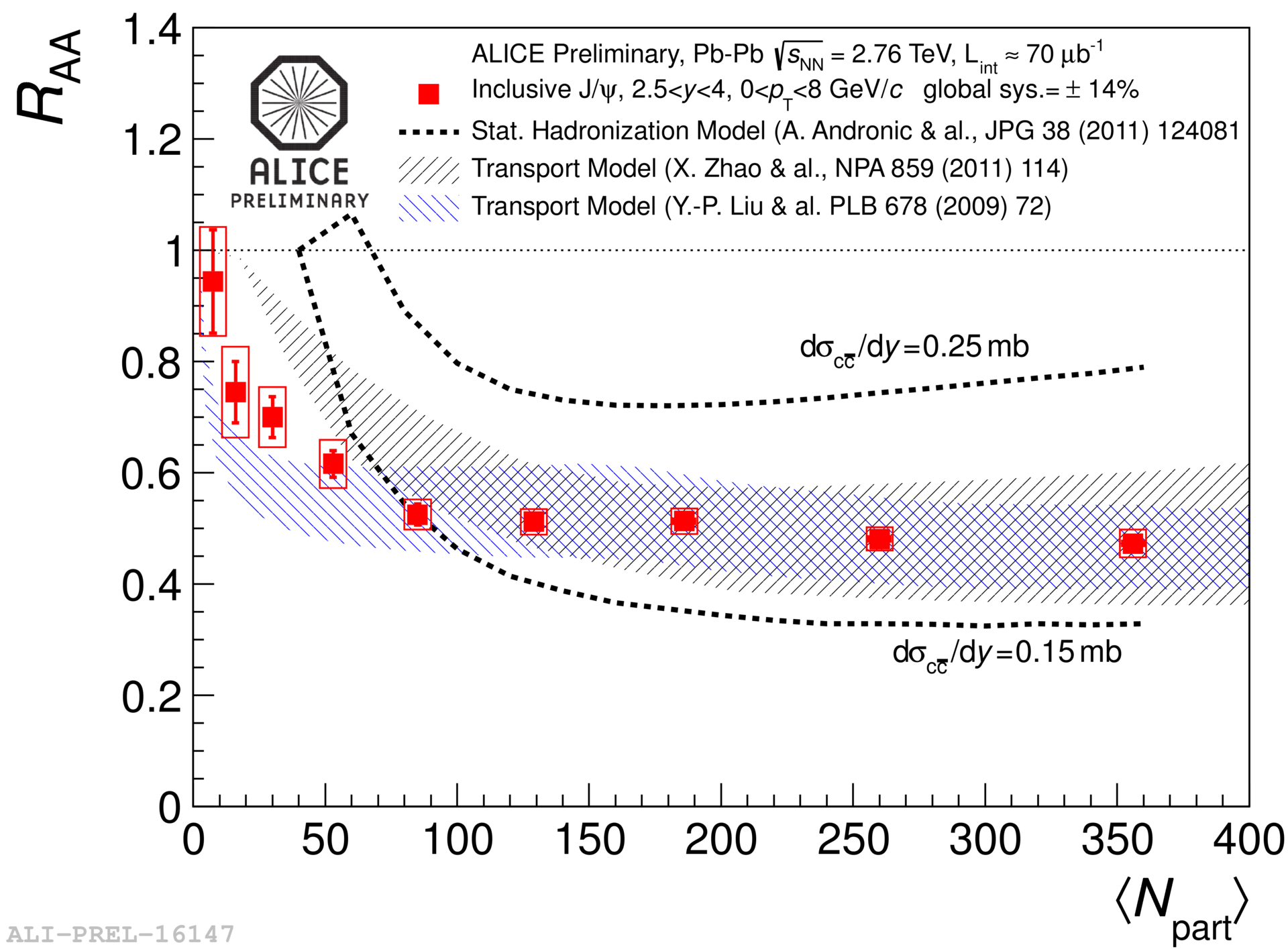}
\includegraphics[width=0.49\linewidth,keepaspectratio]{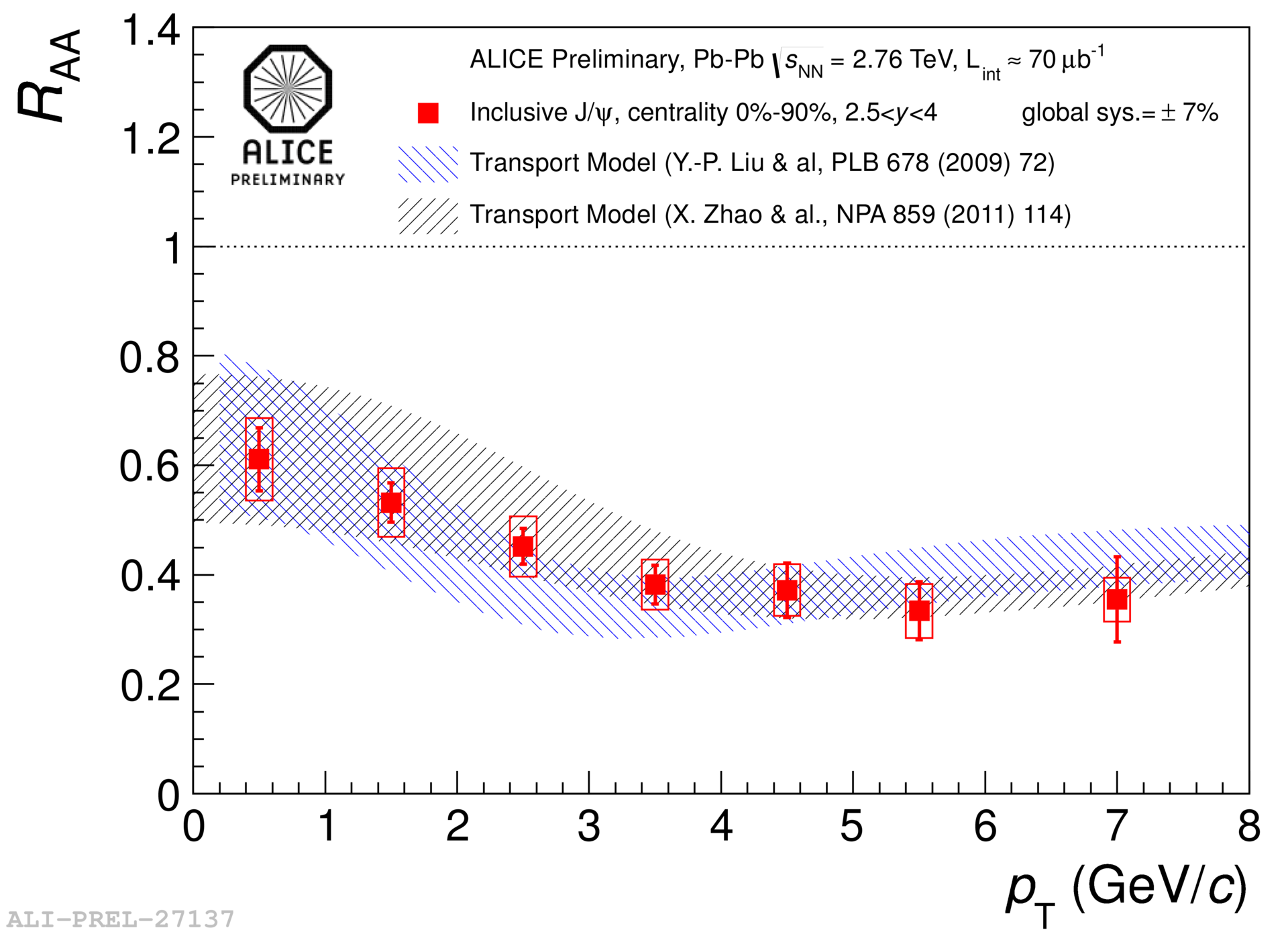}
\caption{(Color online) \label{fig:raamodels} Centrality and \pt\ dependence of the inclusive \jpsi\ \Raa\ measured in Pb-Pb collisions at $\sqrt{s_{\mathrm{NN}}} = 2.76$ TeV  at forward rapidity compared to the predictions by  Statistical Hadronization Model~\cite{Andronic:2011yq},  
Transport Model I~\cite{Zhao:2011cv} and II~\cite{Liu:2009nb}, see text for details.}
\end{center}
\end{figure}

In Fig.~\ref{fig:raamodels}, our  \jpsi\ \Raa\ measurement is compared with theoretical models that all include  a \jpsi\ regeneration component from  deconfined charm quarks in the medium. 
The Statistical Hadronization Model~\cite{BraunMunzinger:2000px,Andronic:2011yq}  assumes deconfinement and a thermal equilibration of the bulk of the  c{$\bar{\rm{c}}$} pairs. 
Then, charmonium production occurs only at phase boundary by statistical hadronization of  charm quarks.  
The prediction is given for two values of  $ {\rm d}\sigma_{\rm{c}\bar{\rm{c}}}/{\rm d}y$ since no measurements are available  at this rapidity  for Pb-Pb collisions. 
The two transport  model results~\cite{Zhao:2011cv,Liu:2009nb}  presented in the same figure differ mostly in the rate  equation controlling the \jpsi\ dissociation and regeneration.  
Both are shown as a band that connects  the results obtained with (lower limit) and without (higher limit) shadowing and can be interpreted as the uncertainty of the prediction. 
The model from Zhao \& al. implements a simple shadowing estimate leading to a 30\% suppression in most central Pb-Pb collisions. 
The charm cross-section d$\sigma_{c\bar{c}}/$d$y$  at $y = 3.25$ is  $ \approx 0.5$ mb and the \jpsi\ from beauty hadrons is estimated at 10\% and no quenching is assumed.
The model from Liu \& al. takes the  shadowing from EKS98 and uses a  smaller charm cross-section d$\sigma_{c\bar{c}}/$d$y \approx 0.38$ mb.
The \jpsi\ from beauty hadrons is estimated at 10\% and b quenching is fixed at \Raa(b)= 0.4 for all  the \pt\ range. 
In both transport models, the amount of regenerated \jpsi\ in the most central  collisions contributes to about 50\% of the production yield, the rest being from initial production.
We can see on the left side of Fig.~\ref{fig:raamodels} that all three models reproduce correctly the centrality dependence of the  forward \jpsi\ \Raa\  for  \Npart\ $ > 70 $. A similar observation   can be made  for the mid rapidity results~\cite{Wiechula:hp2012}.
 The \pt\  dependence of the  forward \jpsi\ \Raa\ is also successfully reproduced by the transport models, as shown on Fig.~\ref{fig:raamodels} right side. 
 In addition, both models predict that a large fraction of \jpsi\ from  regeneration have a \pt\ below $\approx$ 3.5 \gevc .

\section{Elliptic flow in Pb-Pb collisions  $\sqrt{s_{\mathrm{NN}} } = $ 2.76 TeV}
\label{sec:v2results}

\begin{figure}
\begin{center}
\includegraphics[width=0.49\linewidth,keepaspectratio]{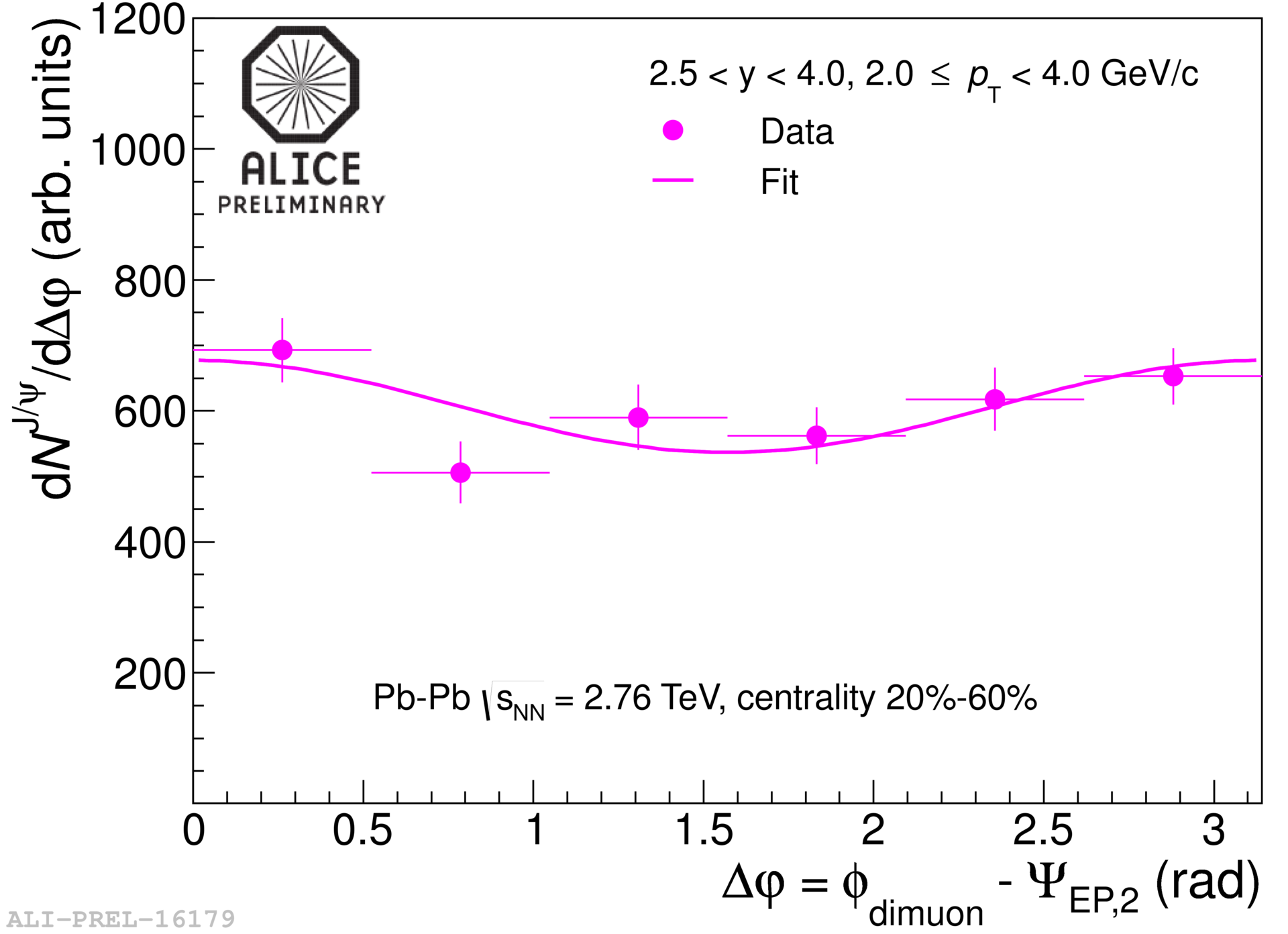}
\includegraphics[width=0.49\linewidth,keepaspectratio]{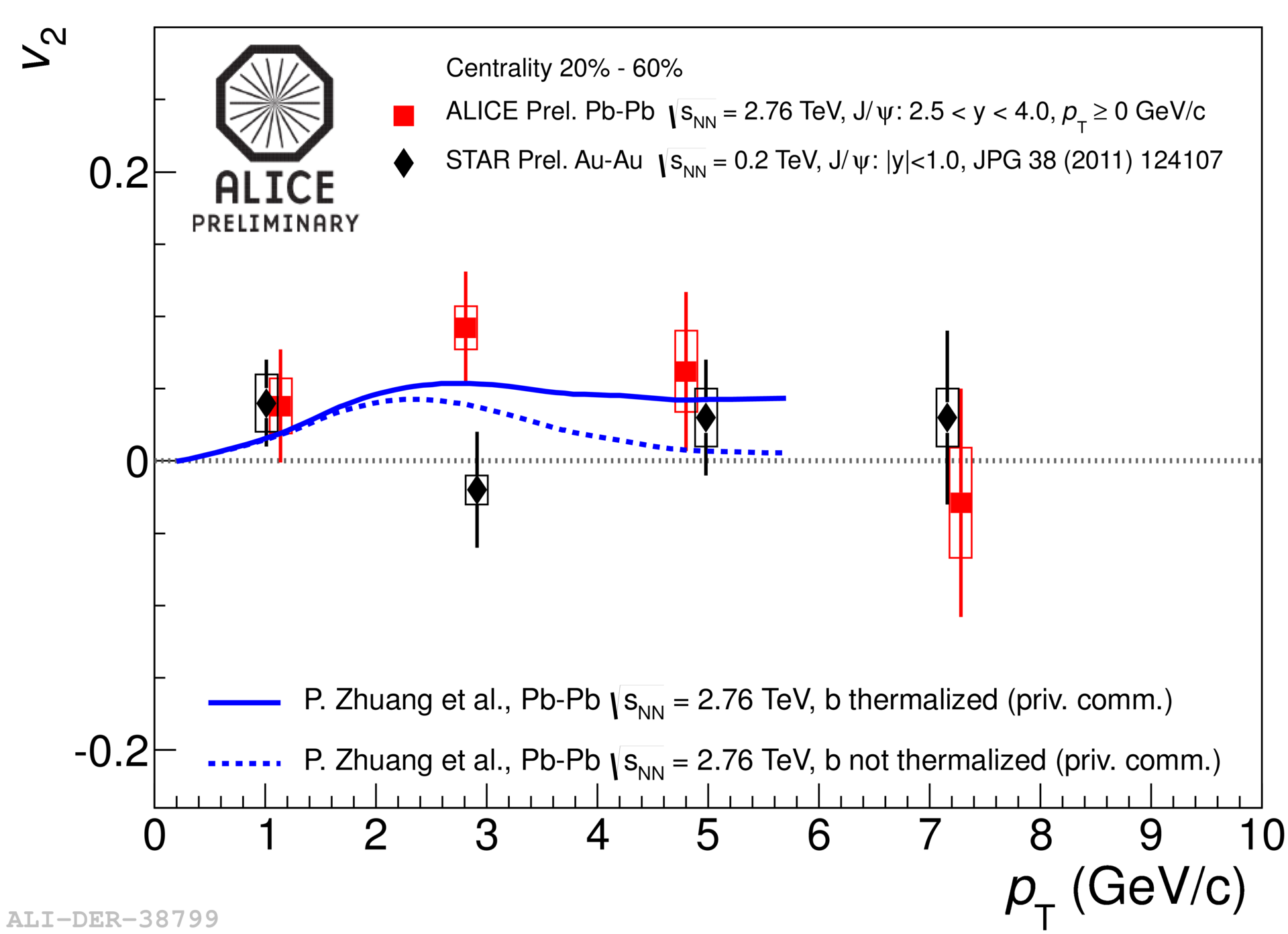}
\caption{(Color online) \label{fig:jpsiv2} Example of the \jpsi\ $v_{2}$ signal extraction in $\Delta \varphi$ bins (left).  
Inclusive \jpsi\ $v_{2}$ measured in the 20\%--60\% centrality range for  Pb-Pb collisions at $\sqrt{s_{\mathrm{NN}}} = 2.76$ TeV (right side) 
compared to the STAR measurement and to the prediction from a parton transport model.}
\end{center}
\end{figure}

The elliptic flow of inclusive \jpsi\ has been measured as a function of the transverse momentum of the \jpsi .
For this measurement, the reaction plane has been determined with the VZERO-A detector. 
The large rapidity gap between the \jpsi\ acceptance and the VZEROA detector minimizes the influence of non-flow effects. 
One the left side of Fig.~\ref{fig:jpsiv2}, an example of the  $v_{2}$ signal extraction is given; one can clearly see the cosine shape of the measured \jpsi\ signal in the 6 $\Delta \varphi$ bins,  where $\Delta \varphi$ is the difference between the azimuthal angle of the \jpsi\  and the angle of the reaction plane. 
Further analysis details can be found 
in~\cite{Massacrier:hp2012}.
Figure~\ref{fig:jpsiv2} shows, on the right side, the first measurement of  \jpsi\  elliptic flow at the LHC. 
The \jpsi\  $v_{2}$ is given as a function of \pt\ in the 20\%--60\% centrality range.
A non-null \jpsi\  $v_{2}$ seems to be present at intermediate \pt\, and would tend to vanish at low and high \pt .
Uncertainties are still too large to draw definitive conclusions, nevertheless we have a non-zero  $v_{2}$ signal  for \jpsi\ with \pt\  between 2 to 4 \gevc\  with 2.2 $\sigma$ significance. 
At lower energy, the \jpsi\  elliptic flow was  measured by STAR  and appear to be consistent with zero at \pt\ $ <$  10 \gevc\  in 20\%--60\% centrality range, 
whereas charged hadrons and  $\phi$ exhibit a rather strong flow in this same kinematic domain~\cite{Tang:2011kr}.
Model prediction (private communication) for ALICE was provided by the authors of~\cite{Liu:2009nb}  and is shown on  Fig.~\ref{fig:jpsiv2}. 
The full line assumes a thermalization of the beauty quarks in the medium, for which there is no evidence so far,  and should be considered as an upper limit. 
The dashed line, a more realistic prediction in which beauty quarks are not thermalized, shows indeed a non-zero  \jpsi\  $v_{2}$,  which matches qualitatively our data.  
It is important to add that this model reproduces successfully  the ALICE \jpsi\ \Raa measurement.

\section{Conclusion}
\label{sec:conclusion}

Quarkonia production in ALICE in pp and Pb-Pb collisions at $\sqrt{s_{\mathrm{NN}}}$  = 2.76 and 7 TeV  has been presented.  
In pp collisions, the \pt,  \y\ and multiplicity dependence of \jpsi\ production,  \jpsi\ polarization and  non-prompt \jpsi\ have been studied.
Results have shown good agreement or complementarity with other LHC results.   
All these measurements provide  stringent constraints to  model predictions.  
In Pb-Pb collisions, the \jpsi\ nuclear modification factor was studied as a function of centrality, \pt\ and \y. 
The \jpsi\ \Raa\  dependence on the number of participant nucleons is flat  and centrality integrated values are large at mid and forward rapidity ($\approx 0.7-0.5$). 
This result is clearly different from the ones seen at lower energies (e.g. RHIC and SPS).  
The rapidity dependence of the \jpsi\ \Raa\ shows that suppression at large rapidity ( 2.5 $<$ \y\ $<$ 4 ) is beyond the one that could be expected from shadowing only predictions. 
The \jpsi\ \Raa\  is large at low \pt\ and then decreases with increasing \pt .  
The trends observed  in the data as a function of  centrality and \pt\ can be reproduced by models  based on deconfinement followed by charm recombination. 
In these models, \jpsi\ from recombination  mostly occur at low \pt\  and account for half of the produced \jpsi\ in the most central collisions. 
Finally, we have presented the \jpsi\ elliptic flow in  semi-central Pb-Pb collisions.  
For the first time, a non-zero \jpsi\ $v_{2}$ is observed in the intermediate \pt\ range.
We have now  accumulated hints  that the \jpsi\ production  in Pb-Pb collisions at LHC energy may be governed, for an important part, by charm quarks recombination processes. 
In order to confirm this observation, the shadowing must be measured and constrained  since it remains unknown at LHC energies and this will be addressed by a pPb  run scheduled at the beginning of 2013. 
One should add here that the uncertainties in \jpsi\ \Raa\ results depend directly on the pp reference data; thus it is crucial to collect a large amount of pp collisions at the same collision energy that of Pb-Pb sample, in order  to have precise measurements of  \jpsi\ and charm differential cross section and \jpsi\ polarization.

\bibliographystyle{elsarticle-num}
\bibliography{suire_HP2012}
\end{document}